\documentclass[aps,prl,groupedaddress]{revtex4}
\usepackage{epsf}
\usepackage{graphicx}        
\input epsf

\def\ll{\label}
\def\re{\ref}

\def\r1{(\ref{$1})}

\def\ba{\begin{array}{c}}

\def\ea{\end{array}}

\def\De{\Delta}
\def\de{\delta}

\def\l{\left}
\def\l({\left(}
\def\r){\right)}
\def\r{\right}

\def\la{\lambda}

 \def\be{\begin{equation}}
\def\bc{\begin{center}}
\def\ec{\end{center}}
\def\bit{\begin{itemize}}
\def\eit{\end{itemize}}
\def\ee{\end{equation}}
\def\ed{\end{document}}
\def\bea{\begin{eqnarray}}
\def\eea{\end{eqnarray}}
\def\efr{\end{flushright}}

\begin{document}
\title{{ Unraveling hidden hierarchies and   dual structures   
in an integrable field model
 }}
\author{ Anjan Kundu\\
 Theory Division,
Saha Institute of Nuclear Physics\\
 Calcutta, INDIA\\
{anjan.kundu@saha.ac.in}}


\begin{abstract} An integrable field theory, due to  
path-independence on the space-time plain, should yield
 together with  infinite sets of
 conserved charges also similar dual charges,
determining the boundary and defect contributions. 
On the example of 
the nonlinear Schr\"odinger equation we unravel hidden hierarchies and 
 dual structures
and show the complete integrability through
 a novel Yang-Baxter equation at the classical and quantum level
with exact solution.
\end{abstract}

 \maketitle

The N\"other's theorem (NT)  connects the symmetries of a model to its
     invariant quantities. 
 In a mechanical system with finite degrees of freedom
such quantities are the first integrals conserved in time. 
For the integrability and exact solvability of a system
the number of  conserved charges should match with 
its degrees of freedom \cite{Laksh}.
This definition is usually generalized straitforwardly to the field theory
(FT) defining  the integrability of a field model through 
 an infinite set of charges  conserved in time.
The NT is also tuned to fit this  definition, stating that
the model should  have infinite number of symmetries with 
one to one correspondence to the conserved charges \cite{Laksh}.  
However, in reality neither the  integrable FT nor the
 NT insists on such   inherently noncovariant properties,
which  are set  perhaps  by  tradition and not by necessity. 
In fact, the NT relates each symmetry of a field model
 to an  integrated  quantity invariant 
along any chosen direction in the space-time manifold \cite{bogolub} and  
similarly an integrable FT,  based on the 
 flatness condition involving both  of its  Lax operators,
allows  a path-independent formulation
on the $x,t $ plain with no emphasis on the 
direction of evolution \cite{FadHam}. However, as mentioned
above,
the conventional approach in  integrable FT focuses mostly on 
   space configurations at a fixed time
 generating  an infinite set of charges   conserved in time,
 through the   space-Lax operator, which  
acts as an infinitesimal shift operator (ISO) along the x-direction. 
This traditional formulation however, as we show, 
 works well  only
for  models  with periodic or vanishing 
boundary condition (BC) and with no     internal defect, 
 but fails for 
more general BC and in the presence of defect points, where the
 usual charges  no
longer remain conserved  \cite {BCFokas,defect}.
In dealing with  the 
 time-dependent boundary conditions, relevant in real
experiments \cite{exp}, 
the  traditional scattering
data  along the space direction is  found also to be insufficient, 
 prompting  the inclusion
  of scattering processes in both  space and 
  time directions
\cite{BCFokas}. These explicit  conflicts  with the  established 
FT force us to
rethink  about the limitations  of the  accepted theory  and  propose
a  dual approach to complement
 the existing results by  unraveling  a wealth of 
hidden structures in the
integrable FT. In particular, working  
 on the example of the nonlinear Schr\"odinger (NLS)
  equation we obtain  an infinite set of dual charges
conserved in space, generated by the 
time-Lax operator acting as  an ISO 
 along the t-direction.
This addresses a rather philosophical question, asking how many
 infinite 
 sets of conserved  charges one  needs for the integrability of a FT,
having  infinite degrees of freedom. By  finding  
  an additional   set of dual charges $ J_k^{(m)}, \ m=2,\ k=1,2,\dots$,
 we could enlarge
 the known set of 
conserved charges $c^{(n)}_k, \ n=1, \ k=1,2,\dots$ 
 for the NLS model by another countable infinity.
More encouragingly, we find that, the contributions from  general BC 
and the defect point can be  given explicitly by these  dual charges,
 which are missed in traditional
 treatments.
 Extending 
to higher Lax pairs one could obtain further  hidden charges with $n=2,3,\ldots $
along with their duals with $m=1,3,4,\dots $, 
generating newer integrable hierarchies covering all possible cases (see
Fig. 1), which are  ignored surprizingly  in the
literature.  

  The   hardest task however is to show the complete 
integrability and exact solvability of the system 
 by  establishing the
 mutual commutativity of   dual charges, in analogy
with that for the usual charges,
 shown  by the celebrated  Yang-Baxter
equation (YBE) linked to the  space-Lax operator 
 and solved by the well known Bethe ansatz (BA) \cite{Fadrev}.
 For this we conjecture  
a    {\it fixed space} Poisson bracket (PB) 
and the related CR for the NLS system, which   could derive 
a novel hierarchy of   integrable equations from the  dual  charges  
 and   at the same time construct 
 a dual  YBE  linked    to the time-Lax operator and
solve the model exactly by  the BA.

The integrable NLS equation 
$ \ i q_t=q_{xx}+2(q^*q)q,/ $
 for the complex field $q(x,t) $
together with its complex conjugate  can be obtained 
 from the 
compatibility of  linear Lax equations  
\be \Phi_x(\la) =U_1(\la) \Phi(\la ) ,\ 
\Phi_t(\la) =V_2 (\la) \Phi( \la ) , \ \Phi ( \la ) 
=(\phi_1 ( \la ), \phi_2 ( \la )), \ll{LaxEq} \ee 
involving  space and  time Lax operators $U_n(\la)$ and 
$V_m(\la), $  with $n=1 $ for the simplest case with space-variable 
$x=x_1 $ and  $m=2 $ for the next higher time variable $t=t_2 $.
 In the conventional approach the generating 
function $\ln \phi_1(x, \la )=i\int_{-\infty }^x dx' \ \rho
(x', \la)$ for the densities $
 \rho
(x, \la)=\sum_k \rho_k (x) (2\la ) ^{-k}
 $ of  charges is considered for vanishing BC  using the simplest
 space-Lax operator $U_1(\la) $. Therefore from the 
 first Lax equation in (\re{LaxEq}) one can derive
  the Riccati equation for $  \frac {\phi_2}
{\phi_1}\equiv  \Gamma (x, \la)= \sum_k \Gamma_k (x) (2\la ) ^{-k}, \ $
solving which one obtains   infinite number  of conserved   charges
 $c_k=\int_{- \infty}^{+ \infty } dx 
\ q \Gamma_k(x) , \ k=1,2,\ldots $, 
 well known for the NLS model  \cite{FadHam}.
The mutual commutativity of the charges
$[c_k,c_l]=0 \ ,k \neq l  $ is proved by the classical and  quantum
YBE involving Lax operator $U_1(\la) $ and the rational $R$-matrix,
 while NLS equation can be obtained 
 from  Hamiltonian  $H=c_3 $  using the usual PB  with canonical momentum $q^* $
\cite{FadHam}.
The known NLS hierarchy is obtained from higher charges
$H=c_{m+1} $, compatible with the flatness condition of the Lax pairs $(U_1,V_m), 
\ m=3,4, \dots \ $ for { time} $t=t_{m} $ (see Fig. 1). 

For exploring  the dual objects we follow a path parallel to the
conventional one, replacing however    $x $ by  
 $t $ and  $U_1(\la) $ by the { time}-Lax operator  
$V_2(\la)$ for the NLS equation.
This  results from the 
second Lax equation in (\re{LaxEq}), describing the scattering process along
the time-direction at  fixed  $x $,
  a generating function
$\ln \phi_1(t, \la )=i\int_{-\infty  }^{\infty} dt' \ j (t', \la)$
 for  densities $
 j(t, \la)=\sum_k j_k (t) (2\la ) ^{-k}, $  of the dual charges $J_k$,
 as $  j
(t, \la)=V_{11}+ V_{12} \Gamma (t, \la), $ 
involving   the    time-Lax operator and  
$\Gamma (t, \la)=\sum_k \tilde \Gamma_k (2 \la)^{-k}.$ 
 Solving  the associated  { time}-Riccati equation 
 recursively  for $\Gamma_k $ we can
 construct  an infinite set of dual charges \ 
$J_k = 
\int_{-\infty}^{\infty } dt \  j_k
(t) ,  \ k=1,2,\ldots , $
 for the NLS equation with vanishing BC on the time-axis
(see Appendix). Similar idea was used earlier for
 obtaining nonlocal charges in the
sine-Gordon field model \cite{Bull81}.  

 Importantly, for  considering  models beyond   periodic
 or vanishing BC in the space interval $[x_0,x_1], $  or
 the models with  a defect at  point $x_d $, 
the traditional  treatment engaging  only the 
space-Lax operator   becomes insufficient
and the inclusion  of the time-Lax operator essential. 
One finds that,  the actual conserved charges 
are not the traditional ones mentioned above, but 
more general ones as
\be C_k= c_k+  I^{BC}_k, \ \   I^{BC}_k=
J_k(x_0,t)-J_k(x_1,t), \ k=1,2,\ldots
,\ll{CcJ} \ee
with $c_k $ as the traditional charges, 
while  $I^{BC}_k $ given through dual charges $J_k $ 
is the  boundary contribution,  or
similarly
\be C_k=  c^-_k + c^+_k + I^{def}_k, \ \   
 I^{def}_k= J_k(x_d^+)-J_k(x_d^-),
\  k =1,2, \ldots \ll{Ckt} \ee
with vanishing BC but with  a defect contribution  $I^{def}_k $ given
 again through  $J_k $ at 
   $
x_d^\pm= \
(x_d\pm \epsilon)_{|\epsilon \to 0} $, where  $ c^\pm_k $  
  are  usual   charges 
 for the fields confined to the 
 left (right)  from the defect point.
It is remarkable that    the dual charges $J_k $, which are
 derived here  but  traditionally ignored, 
determine actually the boundary and 
defect contributions.

Switching over to the dual picture  we can show similarly
that, in more general case  of time-interval $[0,T] $ at fixed
space $x $,   the dual charges  conserved in space generalise to     
\be 
\tilde J_k=J_k +T_k^{BC}, \ \  T_k^{BC}=c_k(0)-c_k(T), \ k=1,2,\ldots,
\ll{tlJcJ} \ee
with time-boundary contribution $T_k^{BC} $  and in  models with 
 {\it  time}-defect 
at moment $t=t_d $ to   
\be 
\tilde J_k=J^+_k +J^-_k +T_k^{def}, \ \ T_k^{def}=  c_k(t_d^+) -
c_k(t_d^-) , \ k=1,2,\ldots,
\ll{tldef} \ee
 with time-defect contribution $T_k^{def}, $
where    $J^\mp_k $ are dual charges in the 
semi-infinite time intervals { before and after} 
  the defect-moment $t_d $.

 However, it is natural to  ask: 
  how far   we can stretch this concept of
 duality. Can 
  dual charges $J_k , \ k=1,2,\ldots, $
 generate a new hierarchy of integrable equations
 including  the NLS equation, mimicking the standard approach but 
  using some unusual canonical bracket?
Defining the dual canonical momentum as
$\tilde p=\frac {\de L} {\delta q_x}$ we derive from the NLS
Lagrangian $\tilde p= q_x^*, \ \tilde p^*= q_x  $ and conjecture
 an   {equal-space} canonical bracket 
$ \{ q(x,t), \tilde p(x,t^{'})\}=\de (t-t^{'}),\ \   
\{ q^*(x,t), \tilde p^*(x,t^{'})\}= \de (t-t^{'}),\ \
\{ q(t), q^*(t^{'})\}=\{ \tilde p(t), \tilde p^*(t^{'})\}=0.
 $
Using this dual PB and taking    $J_2=H $ 
 as 
the   Hamiltonian we derive
 $\tilde p^*_{x}=
\{ \tilde p^*,  J_2\}= iq_t- 2|q|^2q$, 
which by putting  $\tilde p^*\equiv q_x  $ would  yield  the same   NLS equation
associated with the  Lax pair  $(U_1,V_2) $. 
From next higher order   however our dual PB 
 starts  producing new   equations, e.g. for $H=J_3 $ one gets a linear equation 
$q_{t}=q_{x_2} $ related to the pair   $(U_2,V_2) $ and    
for $H=J_4 $, an intriguing integrable  equation
\be
iq_{xt}=q_{x_3}+2(q^*_{x}q-q^*q_{x})
 \ll{dlNLS2}, \ee
 associated with the Lax pair $(U_3,V_2) $ 
  and so on, which generates  a novel integrable  
   hierarchy represented by the flatness condition of the
infinite series of Lax pairs $ (U_k,V_2), \ k=1,2, \ldots$ as a dual  to the 
 NLS hierarchy (see Fig. 1).

 Encoraged by the success with our conjecture
 we set a harder task to establish
 the complete integrability of the system by showing 
 mutual independence of  dual charges 
$[J_k,J_l]=0 , \  k \neq l$ through a  novel dual Yang-Baxter  relation both in  classical
 and   quantum
 cases, where  the  space-Lax operator $U_1(\la) $, appearing in 
the  standard YBE  for the NLS model \cite{Fadrev} is replaced
by  its time-Lax operator
 $V_2(\la) $, 
while the 
classical $r(\la- \mu ) $ and the quantum  $R(\la- \mu ) $ matrices remain
the same  well known rational matrices (see Appendix).

However,  proving the validity of these  dual YBEs 
becomes exceedingly difficult,
   due to much more involved structure 
 of  $V_2(\la) $ for the NLS model  compared to 
 that of $U_1(\la), $ . 
Consequently, 
 in place of just {\it two}
 nontrivial relations appearing in  known  YBEs 
  \cite{FadHam,Fadrev},   
 {\it ten} such  equations
 arise  in   the 
 dual YBEs,
 apart from the ultralocal conditions.   
 However miraculously the dual PB  we propose  
 solves all  classical  relations   
 exactly,
 proving   the dual  YBE
for the classical NLS model. 
 More significantly,  for the  quantum NLS model     
  the corresponding equal-space
commutation relations (CR)  can be  given
 for the field operators   discretize along the 
time axis as
$ \ [q^j(x),{q^{*}_{x}}^k(x)]
={\hbar} {\Delta_{(t)}}^{-1}\de_{jk}, \ \  [q^j(x),{q^{*}}^k(x)]=0
, \ $
$\Delta_{(t)} $ being the lattice constant in discrete time.
 The lattice regularized $V_2^j $ operator 
together with  the rational quantum $R $-matrix and 
 the proposed CR,
   solve fortunately all  nontrivial 
commutation relations   
in the dual quantum  YBE, 
exactly 
 upto  order
  $O(\Delta_{(t)} ) $, which is however   sufficient for the 
QFT model of NLS we are interested in (see Appendix).

Recall that the YBE solutions are the key for exactly solving the 
eigenvalue problem (EVP) for   quantum charges \cite{Fadrev}. 
Using the same idea we  can  formulate an algebraic Bethe
ansatz (ABA) for the exact EVP solution of  the dual quantum charges,
by constructing  a dual monodromy matrix $S(\la)=\prod_j V^j(\la) $ 
 with  the {\it creation} and {\it annihilation } operators defined
through its  operator elements as  $B(\la )= S_{12}(\la)$ and $C(\la )=
S_{21}(\la)$ respectively, while    
 its diagonal elements: $\tau(\la)=tr S(\la) $ 
with $\ln \tau(\la)=\sum_k J_k(2 \la)^{-k} , $ are related to  
   dual quantum  charges $J_k, \ k=1,2,\ldots $. The 
 EVP: $\tau(\la)|N>=\Lambda(\la)|N> $  can be solved exactly in perfect analogy
with  the standard  ABA  formulation \cite{Fadrev}, 
 since the $R $-matrix contribution from our dual YBE remains the
same. Only  the vacuum expectation of 
$U_1(\la) $ appearing in  eigenvalue $\Lambda(\la) $ in the usual case, 
should be replaced     
  by the corresponding value  for the 
 time-Lax operator $V_2(\la) $ in the dual case.
\begin{figure}[!h]
\qquad \qquad
\includegraphics[width=5.5cm,height=5.5 cm]{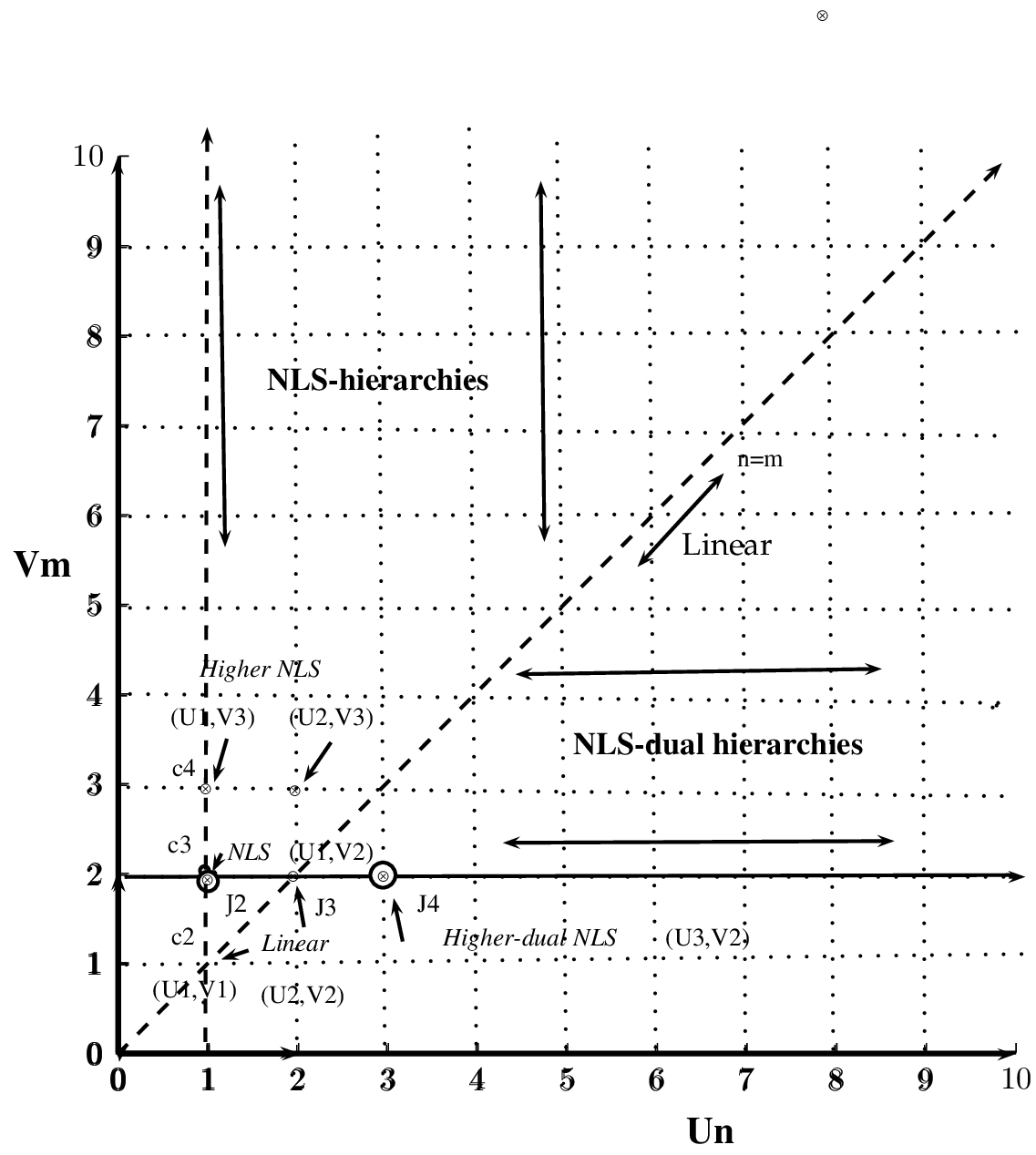}
\caption{Complete NLS hierarchy with Lax pair $(U_n,V_m), \ n,m=1,2,\dots$.
 Vertical lines are the hierarchies with
infinite charges $c^{(n)}_{m+1} $ conserved in time $t_m $  and vorizontal lines
are are the dual-hierarchies with
infinite dual charges $J^{(m)}_{n+1} $ conserved in space $x_n $. The vertical
dashed line corresponds to the known NLS hierarchy, while the horizontal solid line
to the dual hierarchy found here.    
 }  
\end{figure}
We would like to conclude in general that, the  
 NLS field model  posseses 
 rich  hidden  structures and multiple integrable hierarchies beyond
the  known   ones, associated with  
a double-infinite series 
of Lax pairs $(U_n,V_m)  $ with  flatness condition
${U_n}_{t_m}-{V_m}_{x_n}+[U_n,V_m]=0 $, covering all possible cases 
$\ (n,m)=1,2\ldots $ as shown in Fig. 1.  The diagonal line with $n=m  $
represent linear equations since $[U_n,V_n]=0  $ due to $U_n \sim V_n $ , while the vertical 
lines with $ m=1,2\ldots $ for each fixed $n$,  represent integrable hierarchies
with a set of infinite charges $c^{(n)}_{m+1} $ conserved in time $t_m $, of
which only
$n=1 $ corresponds to the known NLS hierarchy.
  The hierarchy with $n=2 $ can be
obtained  by replacing 
  the variables   $t \to {x_2}, \ {x_m} \to {t_m}$ in the explicit form of
the dual hierarchy  found here.
 The horizontal  
lines in Fig. 1   with $ n=1,2\ldots $ 
at fixed $m$,
 represent on the other hand integrable dual  hierarchies  
with a set of infinite dual charges $J^{(m)}_{n+1} $ conserved in space $x_n $.
 $m=1 $ case corresponds to the equations, where the variables in the 
  known NLS hierarchy are replaced as $x \to t_1, \ t_n \to x_n $,
while the case with $m=2 $  corresponds to the dual hierarchy  found above
including (\re{dlNLS2}).
All common  crossing points of  vertical and horizontal lines
must yield the same equations derivable from both   usual  charge and their
duals,
as we have witnessed above  in deriving the  NLS equation 
 associated
with the Lax pair $U_1,V_2 $  at $n=1,m=2 $,  from our dual charge
$J^{(2)}_{2} $  as well as  from the  NLS Hamiltonian $c^{(1)}_{3} $.
     
In summary, using time as dual to the space variable
   we find explicitly an infinite set of dual charges $J^{(2)}_{k}, \ k=1,2,
\ldots $ 
  in addition to the set of usual
  charges $c^{(1)}_{k} $
in an integrable field model,
 on the example of the NLS  equation.  The dual charges generated by the
time-Lax operator $V_2 $ not only determine the defect and the boundary
contributions
 missed in a traditional approach, but also ensure the complete
 integrability
through a dual Yang-Baxter equation, both in  classical and 
 quantum cases with the use of a novel canonical momentum and an equal-space
canonical commutator. Similarly, 
the  contributions of time-boundaries and time-defect
to the general  dual charges are   determined in turn 
by the usual set of  charges,  due to the space-time duality.
It is found that, in general  Lax pairs
$(U_n,V_m) $ with all possible combinations of $(n \in {\bf N}, m \in
{\bf N}) $ 
should generate multiple-infinite sets of integrable hierarchies shown in
Fig. 1, only one
of which (with $(n=1, m \in {\bf N})  $)  is known, while  a few (with $(m=2, n
\in {\bf N}),
\ (m \in {\bf N}
, n=2), \ (m=1, n \in {\bf N})  $ ) are found by us
including the equation (\re{dlNLS2}).
 These are the main results obtained here apart from
the exact solution
 of the NLS quantum field model along the dual direction by algebraic Bethe
 ansatz.
The present investigation  not only 
 exposes the insufficiency in 
 the  established   integrable field theory,
 but  also complements it and  opens up new directions of research
in  integrable classical and quantum  systems,
like  models  with time defects and time boundaries 
as dual to such popular  problems on the space axis,
 exploring the NLS multiple-hierarchy found here for their
 exact  soliton solutions, Hamiltonian structure, YBE etc 
 \cite{kundu2012},         
 new class of
exactly solvable spin chains  along the dual axis 
using different realizations of the quantum time Lax operator,
integrable models on an arbitrary path on a 2D plane
construction of  novel  quantum algebras 
 as dual to the known Hopf algebras in integrable
 systems \cite{kundu99}
   etc.
\vskip .2cm

\noindent {\bf Appendix}\\
\noindent {\bf I. Infinite set of dual charges}
 \smallskip

  Using the matrix  elements  $V_{11}=2\la +|q|^2 ,
V_{12}=V^*_{21}= 2\la q-iq_x$ of the time-Lax operator $V_2(\la) $
for the NLS equation, we can derive its
 dual charges  
$ \  J_k=  
\int_{-\infty}^{\infty } dtj_k, $ \  with $ 
j_k=  (-iq_x \tilde \Gamma_k +q \tilde \Gamma_{k+1}), 
\ k =1,2, \ldots \ , $   
through the  solution $\Gamma_k $ of the time-Riccati equation
$ \  i{\Gamma_k}_t= \Gamma_{k+2}-2| q|^2\Gamma_{k}
 +\sum_{j}(q\Gamma_{j}\Gamma_{k+1-j} - iq_x 
\Gamma_{j}\Gamma_{k-j})
$
with  densities $j_k $ as    
   \ \
$ j_1= i (q^*_xq-q^*q_x), \  j_2=  iq^*_{t}q+ q^*_{x}q_{x} + | q|^4
 , \  j_3= q^*_{t}q_x , \ \    j_4=  
iq^*_{xt}q_{x}+q^*_{t}q_t-
i |q|^2(q^*q_{t}- q^*_tq) 
 -2 |q|^2 \ q^*_xq_{x}+( {q^*}^2q^2_{x}+
{q_x^*}^2q^2
 ), $
\ \ and so on.

\noindent {\bf II. Dual  classical  YBE}
 \smallskip

The  dual YBE  of the NLS model
is  linked to the time-Lax operator $V_2(\la) $ with  the same rational 
 r-matrix as 
\ \ $\  \{ V(\lambda,x,t_1) \otimes _, V(\mu,x,t_2) \}
=[r(\la-\mu), V(\mu,x,t_1)\otimes I+I\otimes V(\mu,x,t_1)  ]
 \de (t_1-t_2), \ $\ \ 
In  this matrix   relation 
 ten equations  remain nontrivial, which however 
    using the dual PB can be  reduced into only 
two  coinciding or related 
groups with values 
 $-i q $ and $ 
-2i (\la +\mu)  $ multiplied  by  $ {\de (t_1-t_2)} $ on both
 sides of
the equation proving the dual YBE.
 
\noindent {\bf III. Dual  quantum  YBE} 
\smallskip

The dual YBE for the NLS quantum field model with  regularized time Lax operator
  $V^j $    
  and  the known rational 
$R$-matrix is
\\ $R(\la-\mu)V^j (\la ) \otimes V^j(\mu ) =
(I \otimes V^j(\mu ))(V^j(\la ) \otimes  I) R(\la-\mu )
, \ j=1,2, \ldots N. $\\
In  the lattice regularization 
$V^j(\la)=I+i\Delta_{(t)} V_2(\la) , \ \Delta_{(t)}| \to 0, $
 the field operators 
$q(x,t), q_x^*(x,t)$ in the known time-Lax operator $ V_2(\la )$ 
are replaced   by  their
  semi-discretize variants $ q^j(x),\ 
 {q^*}^j_x(x) $.
This results to  ten nontrivial  operator equations
containing  terms of order 
 $O(\De _{(t)} )  $ apart from higher order terms.
For example, in element    $(11,12) $: 
$ \  \eta (V^j_{11}(\la)V^j_{12}(\mu ) -V^j_{11}(\mu )V^j_{12}(\la ))+
(\la -\mu)(V^j_{11}(\la)V^j_{12}(\mu ) -V^j_{12}(\mu )V^j_{11}(\la )),
\ $
  the first term is nontrivial: 
  $-2\eta \De _{(t)}  (\la -\mu) q^j $. However
the second term containing  
 $i\De^2 _{(t)}  (\la -\mu) [{q^j}^\dagger q^j,q_x^j]  $, 
  cancels  it  exactly  by  
 using the proposed dual CR.
Similar picture is repeated for other  terms
inducing exact cancellation of   all terms upto order $O(\De _{(t)} )  $.
The   ultralocal condition
$ \   V^j(\la ) \otimes   V^k(\mu )=V^k(\la )
 \otimes  I)(I \otimes V^j(\mu )), \ k \neq l. \ $
follows from the canonicity of the dual CR.

The author is indebted to Prof. P. Mitra for stimulating discussions.

 \end{document}